# Fast and Reconfigurable Packet Classification Engine in FPGA-Based Firewall

Arief Wicaksana[#1], Arif Sasongko[#2]

[#]School of Electrical Engineering and Informatics
Institut Teknologi Bandung, Bandung, Indonesia
[1]ariefgrand@students.itb.ac.id
[2]asasongko@stei.itb.ac.id

*Abstract*— In data communication via internet, security is becoming one of the most influential aspects. One way to support it is by classifying and filtering ethernet packets within network devices. Packet classification is a fundamental task for network devices such as routers, firewalls, and intrusion detection systems. In this paper we present architecture of fast and reconfigurable Packet Classification Engine (PCE). This engine is used in FPGA-based firewall. Our PCE inspects multi-dimensional field of packet header sequentially based on tree-based algorithm. This algorithm simplifies overall system to a lower scale and leads to a more secure system. The PCE works with an adaptation of single cycle processor architecture in the system. Ethernet packet is examined with PCE based on Source IP Address, Destination IP Address, Source Port, Destination Port, and Protocol fields of the packet header. These are basic fields to know whether it is a dangerous or normal packet before inspecting the content. Using implementation of tree-based algorithm in the architecture, firewall rules are rebuilt into 24-bit sub-rules which are read as processor instruction in the inspection process. The inspection process is comparing one sub-rule with input field of header every clock cycle.
The proposed PCE shows 91 MHz clock frequency in Cyclone II EP2C70F896C6 with 13 clocks throughput average from input to output generation. The use of tree-based algorithm simplifies the multidimensional packet inspection and gives us reconfigurable as well as scalable system. The architecture is fast, reliable, and adaptable and also can maximize the advantages of the algorithm very well. Although the PCE has high frequency and little amount of clock, filtering speed of a firewall also depends on the other components, such as packet FIFO buffer. Fast and reliable FIFO buffer is needed to support the PCE. This PCE also is not completed with rule update mechanism yet.
This proposed PCE is tested as a component of FPGA-based firewall to filter Ethernet packet with FPGA DE2 Board using NIOS II platform.

I. INTRODUCTION

In data communication via internet, security is becoming one of the most influential aspects. A system must provide ways to send and receive data and information securely. As Ethernet device is massively used on this communication, therefore Ethernet packets is the most used format of the data within network. The network must be able to classify and filter this packet. Generally, network devices classify packets into two categories, permitted packets and blocked packets. Permitted packets are forwarded to the next step while blocked packets are removed from the line. Many institutions connect to the Internet, guarded by "firewalls" designed to prevent unauthorized access to their private networks [3].

A firewall is a dedicated appliance, or software running on a computer, which inspects network traffic passing through it, and denies or permits passage based on a set of rules. Frankly speaking, a firewall is a boundary network device that resides between a private network and the Internet [7]. Its basic function is to regulate the flow of traffic between computer networks of different trust level [1]. Firewall forms the basic building block of a network security architecture making them indispensable due to growing security concerns brought by the explosive growth of the Internet. A firewall enforces security policy between two networks by selectively permitting or blocking traffic between them [2]. There are two basic approaches [3],[8] for security policy, 'default deny' and 'default allow'. 'Default deny' means packet is always blocked unless it is specified, and 'default allow' is the opposite.

Conventional firewalls operate at the network layer and their operations are based on stateful or non-stateful filtering [1] which is quite reliable to prevent unauthorized access to private network. Software-based firewall has more capabilities in packet-filtering than hardware-based firewalls although it has lower speed than the later. Hardware-based firewall is unable to make more complex decision. This leads to less security and functioning more like a router from the packet filtering point of view.

Packet classification is a fundamental task for network devices such as routers, firewalls, and intrusion detection systems. Packet classification is performed using a packet classifier, also called a policy database, flow classifier, or simply a classifier [4]. A basic firewall decides packet classification from at least 5 header fields namely, Source IP Address, Destination IP Address, Source Port, Destination Port, and Protocol field of an IP packet header.

In this paper we propose a fast and reconfigurable PCE. Classification rules in the firewalls are broken down into sub-rules to decrease amount of bits in data memory and simplify inspection process using tree-based algorithm. Although use of index or encoding in the firewalls rules [2],[4]-[6] can decrease amount of bits, it has certain complexity in the architecture. The complexity is caused by the process of indexing the classification rules. The index has to be saved in memory or it will need certain architecture to convert index into a real data. Our algorithm avoids the use of index and utilizes tree shape to implement steps of inspection. This will lead to simple field checking and less bits in the classification process. Our packet classification engine also adapts single cycle processor architecture to implement header-fields inspection. Each header fields are inspected with broken down rules as sub-rules by the packet classification engine within one cycle. The sub-rules are fetched from memory similar to instruction fetching in a single cycle processor. This architecture makes our firewall is reconfigurable and scalable to suffice different levels of protections. Modification or rules update to increase level of firewalls security and amount of rules are to be done only in the memory part that will control the inspection process in the engine, which is fast, simple, and reliable to provide packet inspection based on the rules implemented. Our proposed packet classification engine has capabilities to overcome most of hardware-based firewall weakness which is unable to make complex decision.

This paper is organized as follows: Section II presents the algorithm used in packet classification to implement the firewall rules, Tree-based Algorithm. Section III shows the architecture of packet classification engine. Section IV presents the conclusion of this paper. Section V shows the feature of our proposed packet classification engine in the FPGA-based firewall and the last section presents references of this paper.

## II. TREE-BASED ALGORITHM

Rules implemented in the PCE are configured by the designer of firewall. The functionality of PCE follows all the statements in the rules. In our proposed PCE, we use 'default deny' as the basic strategy to prevent all unspecified packet through the network. The proposed PCE uses algorithm based on the reduction of complexity in classification rules and its modification into lowers scale of rules. Table I shows example of rules implemented in a firewall to do packet inspection. A rule of firewall is made from combination of some fields from packet header. In the table, combination of 5 packet header fields forms one rule. In other words, there is 5 dimension equation to generate one value if all the specification is fulfilled, "allow" or "deny".

Rules simplification is approached using an algorithm which uses tree as a basic form, we called it Tree-based Algorithm. Just like a tree, each process consists of root and leaves and committed sequentially from root to the leaves. Root of one process is the most important step and should be done first. Root step also means that the step has fewer options and potentially simplifies the works if it is done immediately. Sequences of packet inspection based on header fields in PCE are shown in Figure 1. The inspection is implemented basically with only 5 fields, Source IP Address, Destination IP Address, Source Port, Destination Port, and Protocol field. It is begun with Protocol inspection and lasted with Destination Port. In the figure, Protocol is the root and Destination Port is the leaf. This process is done step by step, where the next step cannot be done until the previous one is finished. This algorithm gives dependencies among each packet inspection which makes relatively more secure system and stable.

TABLE I
EXAMPLE OF FIREWALL RULES

| Source IP Address | Destination IP Address | Source Port | Destination Port | Protocol | Action |
|---|---|---|---|---|---|
| 167.205.3.11 | 167.205.65.32 | 25 | 8080 | TCP | Allow |
| 192.168.*.* | *.*.*.* | 80 | * | TCP | Deny |
| 167.205.65.5 | *.*.*.* | * | * | UDP | Allow |
| *.*.*.* | 134.25.5.2 | >1023 | 80 | TCP | Allow |

For each step in inspection field tree, there are inspection trees described those steps of tree. Fields inspection is broken down into sub-fields inspections. Sub-field is a part of a field which is decided to consist 8 bit of each field. Inspection sub-field tree represents various level of sub-field in a field inspected. Figure 2 shows inspection sub-field tree for Protocol field in the main tree. Protocol is the root of the tree and options of protocol are the leaves. Options of protocol are various protocols specified in the firewall rules. For every option, an action is specified by PCE to handle the process, block the packet, forward the packet, or jump to next step of the process. Protocol field consists of 8 bit so it will have 1 level of tree.

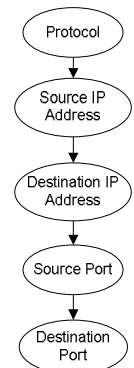

Fig 1. Inspection Sequence Based On Header Fields

Our proposed PCE compares Protocol field input to the system with various protocol options in the rules and search for the most suitable option. If it satisfies with an option, an action related to the option will be done to the inspection process, forward the packet directly or jump to next field inspection. If it does not satisfy with any option, the packet will be blocked since it has "deny all" strategy default. This "deny all" condition satisfies to all the protocols which are not specified. This algorithm saves time and source to do non-advantage packet inspection.

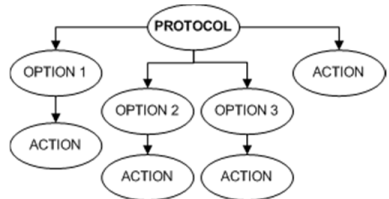

Fig 2. Protocol Inspection Process

Figure 3 shows next field inspection step, Source IP Address Inspection. Source IP Address consists of 23 bit which forms 4 sub-fields, from MSB to LSB. 4 sub-fields in Source IP Address make 4 level of tree. In the figure, we can see from SA_1 to SA_4. In each level, there are options based on rules implemented in the firewall. Sub-field input is compared with sub-field options in the rules to find the most suitable category. And also for each level, there are actions for success and failed inspection. This process is done sequentially until all the level of Source IP Address tree inspected. Destination IP Address field is also inspected with similar mechanism. Destination IP Address inspection tree is shown in Figure 4.

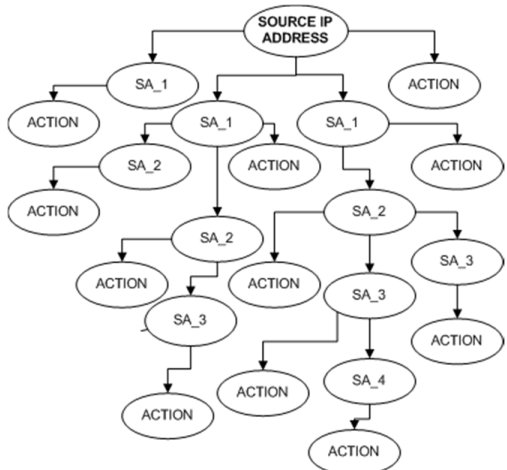

Fig 3. Source IP Address Inspection Process

Next step of the process in the inspection tree is Source Port and Destination Port field inspection. These fields consist of 16 bit which have 2 level of tree. Their inspection trees are shown in Figure 5 and Figure 6. Generally, the process omitted in Source Port and Destination Port inspection is similar with previous step. This is why we believe the tree-based algorithm is reconfigurable and applicable to different amount of steps and dimensions in the rules.

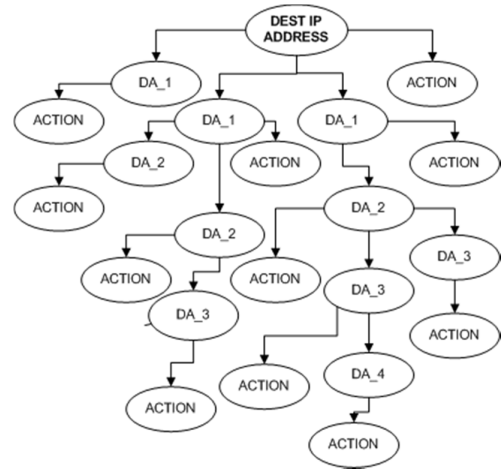

Fig 4. Destination IP Address Inspection Process

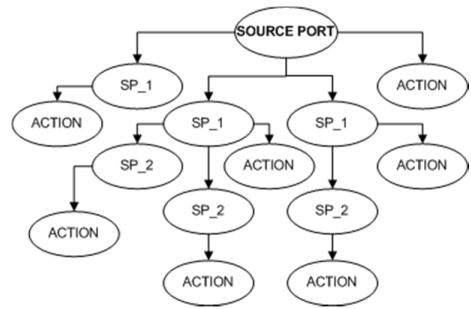

Fig 5. Source Port Inspection Process

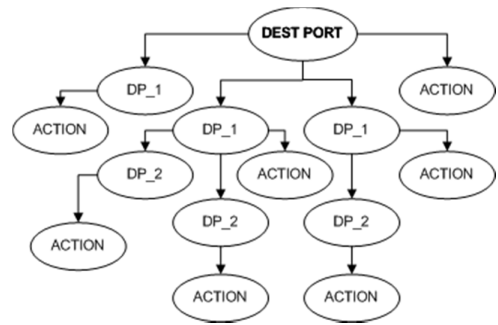

Fig 6. Destination Port Inspection Process

To make the firewall even more secure, one can add various conditional specifications or amount of fields inspected. Tree-based algorithm provides ease to modify the PCE specification without architectural modification by following the procedure explained step by step. Architectural modification also can be done with the system still use the algorithm. Our proposed PCE is made from combination of unrelated systems but support each other. In the next section, we will explain about the architecture of packet classification engine.

## III. ARCHITECTURE OF PACKET CLASSIFICATION ENGINE

This section explains about steps to design Packet Classification Engine architecture. The design is based on specification explained above. PCE implements basic inspection from 5 header packet field in ethernet packet. Those five fields are input to the system. Figure 7 shows Top Level of PCE designed. The inputs to the system are Source IP Address, Destination IP Address, Source Port, Destination Port, Protocol, START, RESET, and CLOCK signal while the outputs are Valid and Forward.

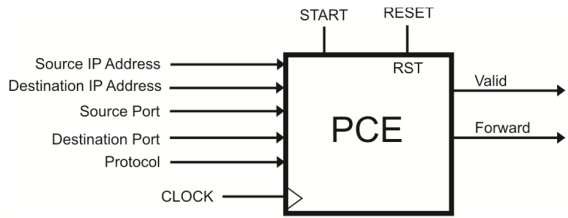

Fig 7. PCE Top Level

START signal is a control signal to start an inspection process by PCE. Every inspection is controlled by this signal to receive field inputs as system inputs and process it. RESET signal cleans register value and outputs, and initialize the system. Output Forward signal shows 1 inspection process is finished and read Valid signal as a result. Valid signal shows fields inspection result and control packet buffer to forward or discard the ethernet packet from the firewall. CLOCK signal is needed to synchronize the system process.

Process in PCE starts from input and ends with output signal generation, shown by High Level State Machine in Figure 8. When the system is turned on, PCE will be in the idle state. In this state, the system waits for start signal input gives logic '1'. When the signal is given, all the field inputs are received and system will move to process_1 state. In this state, all the output is clean. PCE inspects the input fields based on the algorithm. Every one inspection, it moves to process_2 state and checks for inspection status. When the inspection is not finished, it will loop back to process_1. When it is finished, the system moves to stop state. In this state, outputs are generated and the system will move to idle state automatically until the next process.

Based on the algorithm explained above, we need an architecture which can maximized its advantages and works fast and reliable. The architecture must have simple design but also have the capability to facilitate various modifications of firewall rules and specification. The idea of building our PCE architecture comes from the algorithm itself. This PCE takes form of a single cycle processor whose processes are decided by instruction memory. This architecture is most suitable for PCE in a firewall which implements variable rules and specification. Each of 8 bit sub-fields from tree-based algorithm will produce instructions for the processor.

From 5 inspected fields of packet header, we will get 13 sub fields, 4 sub-fields of Source IP Address, 4 sub-fields of Destination IP Address, 2 sub-fields of Source Port, 2 sub-fields of Destination Port, and 1 sub-field of Protocol. With 13 sub-fields inspected by PCE, we need mechanism to decide which sub-field is inspected. The processor will need a multiplexer with 4-bit selector for the data processed. Figure 9 shows the multiplexer used. Before the multiplexer, there is a register to secure the system from unwanted input change.

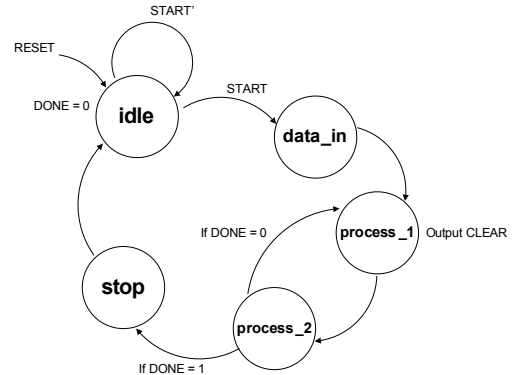

Fig 8. High Level State Machine of PCE

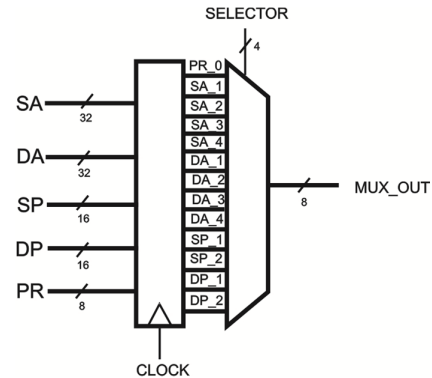

Fig 9. Sub-Fields Multiplexer

The most important component in sub-fields inspection by PCE is comparator. It compares sub-field input with sub-fields in the firewall rules. The comparison process has been explained in the tree-algorithm. Comparator is designed to compare 2 input 8 bit data and give result based on the criteria. These criteria are greater, less, or equal. If the comparison of 2 inputs and the criteria matches, this component will generate logic '1' result, and otherwise if it is not. Figure 10 shows comparator used in the PCE. The criteria is 2 bit input at the component.

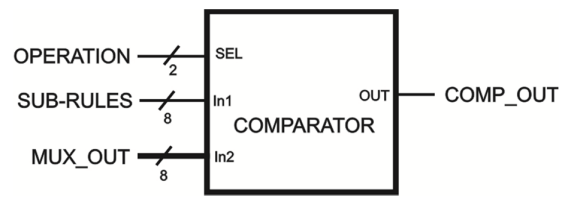

Fig 10. Comparator

The architecture needs a component to control system output at the right time, which is after field inspection process is finished. Logic '1' from the comparator only shows that the comparison is suitable but the process is not over yet. Final compile unit is designed to generate output if the process is finished. Valid and Forward signal in PCE is generated by this component. Command to generate output will be saved in the last part of a process in the accessed rules by the system. Output generation is shows by the action leaves at the end of inspection tree. Final Compile Unit is shown in Figure 11. 1 bit is needed to control output generation.

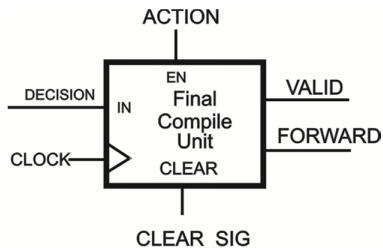

Fig 11. Final Compile Unit

By taking processor form as the architecture, PCE needs instructions to control process undergoing in each cycle. These instructions are saved in instruction memory. In our proposed PCE, the component saving the instructions is called Rules Memory. Each instruction can be accessed by giving address input to the component. Figure 12 shows Rules Memory used in PCE. Because in this research we do not use complex memory and updating mechanism yet, Rules Memory address is consisted only 8 bit. This address will change based on the specification.

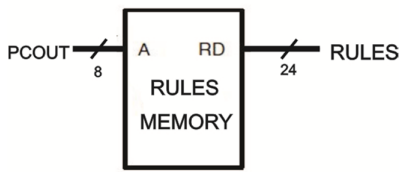

Fig12. Rules Memory

Addressing in rules memory needs a component to continue the process to next address. This component is Program Counter. Figure 13 shows Program Counter used in PCE. Program Counter inputs are previous address, clock signal, reset, and hold. By default, Program Counter start counting from "00000000" address and add 1 to previous address each cycle to the next address, except if there is jumping address procedure. Hold input is used to stop the process if the inspection is finished. It is done so PCE does not have to works while inspections are not needed. Jumping procedure to certain address happens when one step of a process is finished and it continues to certain address. Jumping procedure is controlled by 1 bit jump. Choosing next address between addition result of the previous address or certain address in jumping procedure is done by 2 input multiplexer with 1 bit selector. The selector shows continuation of the process, whether it is caused by jumping procedure or suitable comparison.

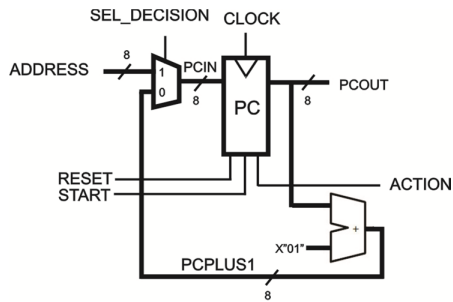

Fig 13. Program Counter

Each 8-bit field from tree-based algorithm is integrated with controller and input bits for each component. The result is processor instruction and saved in the rules memory. This processor instruction is 24 bit data and called sub-rules because of its function controlling inspection process in PCE. Figure 14 shows the configuration of 24-bit sub-rules bit in the rules memory. The data saved in rules memory is shown in Table 2.

| 1-bit JUMP | 4-bit SELECTOR | 2-bit OPERATION | 8-bit HEADER FIELD | 8-bit ADDRESS | 1-bit ACTION |
|---|---|---|---|---|---|
|  |  |  |  |  |  |

Fig 14. Component of 24-bit Sub-Rules

TABLE 2
DATA SAVED IN RULES MEMORY

| Rules Address | Rules Data |
|---|---|
| 00000000 | 000001100000001000001000 |
| 00000001 | 000001100010001001001000 |
| 00000010 | 000001100001100010101000 |
| 00000011 | 000000000000000000000001 |
| etc. | etc. |

From all of the components explained to inspect fields by PCE, complete architecture of PCE can be built. This architecture is shown in Figure 15.

IV. CONCLUSIONS

The proposed packet classification engine shows 91 MHz clock frequency Cyclone II EP2C70F896C6 with 13 clocks throughput average from input to output generation. The use of tree-based algorithm simplifies the multidimensional packet inspection. It gives us reconfigurable as well as scalable system since it is easy to modify or enlarge the system without building a new architecture. The architecture used works fast, reliable, and adaptable with any firewalls modification similar with general purpose processor. Its combination can maximize the advantages of the algorithm very well.

Although the PCE has high frequency and little amount of clock, filtering speed of a firewall also depends on the other components, such as architecture of packet FIFO buffer. This FIFO buffer is used to keep packet data while it is inspected.

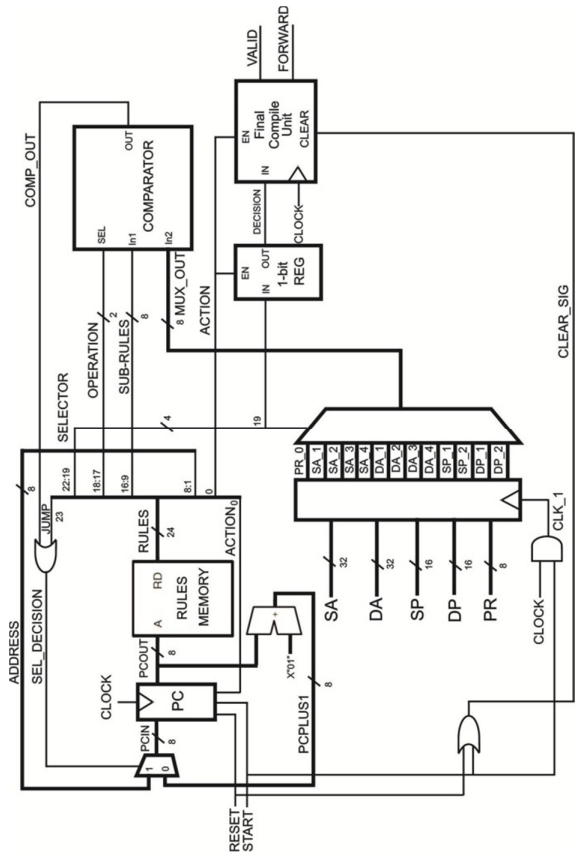

Fig 15. Packet Classification Engine Architecture

Fast and reliable FIFO buffer is needed to support this packet classification engine. Our packet classification engine also is not completed yet. As a complete firewall, it needs rules update mechanism to reach more secure system from its experience. This firewall rules update mechanism is not implemented yet in the PCE.

Packet classification engine is not enough to build an integrated high speed firewall. In future works, we need to design a custom packet FIFO buffer in our system. Rules update mechanism in software-based is also needed to make more reliable system and to handle more complex rules. The packet classification engine architecture also has potential to be optimized since in this paper we still focused in its functionality only.

## V. TESTING

This proposed PCE is tested as a component of FPGA-based firewall to filter Ethernet packet in FPGA DE2 Board. Our PCE is integrated with other components in SOPC system [9]. Figure 16 shows the implementation outline in our FPGA-based firewall. PCE works as one of the FPGA component based on hardware while the other components work based on software in NIOS II Platform. It filters unspecified packet Ethernet in the network.

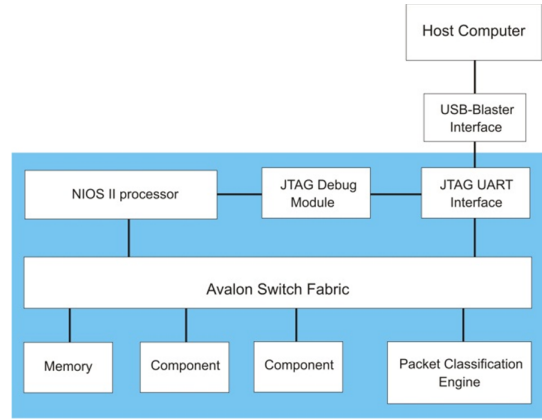

Fig 16. Implementation Outline of PCE in DE2

ACKNOWLEDGMENT

This paper was prepared as a part of research in hardware-based firewall documentation in Electronics Department Institut Teknologi Bandung. The topic of the research was conducted by Mr. Sarwono Sutikno, lecturer and first writer's supervisor.

Other contributions were from School of Electrical Engineering and Informatics and VLSI-RG Research Group to facilitate the research devices. We also thank the reviewers for their helpful comments.